\documentstyle[aps]{revtex}\twocolumn
\begin{document}
\title{The Range of Validity for the Kelvin Force}
\date{PRL {\bf 84}, 2762, (2000)}
\maketitle In a recent Letter~\cite{luo}, Luo, Du
and Huang reported a novel convective instability
driven by a force rarely studied before -- that
exerted by an external magnetic field on a strongly
magnetizable liquid. The associated physics is
surprisingly rich and promises many more interesting
results for the future. Unfortunately, the analysis
starts from a misconception and employs the Kelvin
force outside its range of validity. Since few would
recognize this as a mistake, and since its
consequence in the given experiment is particularly
direct and critical, this is a point well worth
being clarified, and clearly understood.

In the experiment, ferrofluid is exposed to a constant
$\bf B$-field. Yet, since the temperature $T$ and the
density $\rho$ of magnetic particles vary, so does the
magnetic field ${\bf H}={\bf B}/[1+\chi(T,\rho)]$, giving
rise to a finite Kelvin force. With $\chi$ the magnetic
susceptibility and ${\bf M}=\chi{\bf H}$ the
magnetization, this force is given as
\begin{equation}
{\bf f}=M_i\mbox{\boldmath$\nabla$}H_i=(M_iB_i)
\mbox{\boldmath$\nabla$} \frac{1}{1+\chi}
=-\frac{(\chi
B)^2}{(1+\chi)^3}\frac{\mbox{\boldmath$\nabla$}\chi}{\chi}.
\end{equation}
(Summation over the index $i$ is implied.) Eq~(1) may be
derived from the more general Helmholtz force~\cite{LL8},
\begin{equation}\label{87}
{\bf f}= +\mbox{\boldmath$\nabla$}
(\textstyle\frac{1}{2}H^2 \rho\partial\chi/\partial\rho)
-\textstyle\frac{1}{2}H^2\mbox{\boldmath$\nabla$}\chi,
\end{equation}
by considering a dilute ferrofluid, and taking $\chi$ as
proportional to the particle density $\rho$, or
$\rho\partial\chi/\partial\rho=\chi$. Then Eq~(2) clearly
reduces to ${\bf f}=\textstyle\frac{1}{2}
\chi\mbox{\boldmath$\nabla$}
(H^2)=M_i\mbox{\boldmath$\nabla$}H_i$.

All this seems rather convincing, but in fact hides a
pitfall. Closer scrutiny reveals that ${\bf
f}=M_i\mbox{\boldmath$\nabla$}H_i$ is only valid to
linear order in $\chi$. (Except in unconventional systems
of more recent dates, the magnetic susceptibility $\chi$
is usually much smaller than 1, so terms of higher order
in $\chi$ have always been negligible. This may well be
the reason why the confined range of validity of the
Kelvin force has been such a well kept secret.) If true,
the expressions of Eq~(1) merely states that the force
vanishes -- to linear order in $\chi$. No result derived
from Eq~(1) is then trustworthy.

To qualitatively understand this restriction, define a
different susceptibility, ${\bf M}=\tilde\chi{\bf B}$.
With the permeability given as $\mu= 1+\chi
=(1-\tilde\chi)^{-1}$, we have $\tilde\chi=\chi/
(1+\chi)$. Both susceptibilities are clearly physically
equivalent, and we have no a priori reason to prefer
either. Employing ${\rm d}\tilde\chi={\rm
d}\chi/(1+\chi)^2$, we may rewrite Eq~(\ref{87}) as
\begin{equation}\label{87a}
{\bf f}=+\mbox{\boldmath$\nabla$}
(\textstyle\frac{1}{2}B^2\rho_\alpha
\partial\tilde\chi/\partial\rho_\alpha)
-\textstyle\frac{1}{2}B^2\mbox{\boldmath$\nabla$}
\tilde\chi.
\end{equation}
This time, assuming $\tilde\chi$ as proportional to
$\rho$, we obtain
\begin{equation}
{\bf f}=M_i\mbox{\boldmath $\nabla$}B_i,
\end{equation}
a result obviously different from Eq~(1) -- but one that
also vanishes for uniform $B$-fields, so there is no
disagreement to ${\bf f}=M_i\mbox{\boldmath $\nabla$}H_i$
in linear order.

Now, since Eqs~(\ref{87}) and (\ref{87a}) are
algebraically equivalent, the difference must lie between
the two seemingly innocuous assumptions, $\chi$ or
$\tilde\chi\sim\rho$. Reviewing the above derivations, it
is obvious that if one of the two assumptions were {\em
strictly} correct, the other would be wrong, and only the
associated force expression is applicable.

Generically, on the other hand, both $\chi$ and
$\tilde\chi$ are power series of $\rho$. So we are simply
approximating, discarding quadratic and higher order
terms, when we assume that either is linear in $\rho$.
The consistent dilute limit is given when all terms
$\sim\rho^2$ (and higher) are discarded. With
$\chi\sim\rho$, this necessarily implies that we must
also discard all terms $\sim\chi^2$. As a result,
$\tilde\chi=\chi/(1+\chi)\approx\chi$ and
$M_i\mbox{\boldmath $\nabla$}B_i\approx
M_i\mbox{\boldmath $\nabla$}H_i$. We conclude: {\em The
Kelvin force is valid to linear order in the density
$\rho$ and the susceptibility $\chi$ (or magnetization
$M_i$). Especially, both $M_i\mbox{\boldmath
$\nabla$}B_i$ and $M_i\mbox{\boldmath $\nabla$}H_i$ are
valid expressions for the Kelvin force.}

The force in the experiment of~\cite{luo} is of course
finite. A proper, quantitative evaluation is given by
including terms of higher order in $\rho$. As a
first-step, we consider the next order terms:
\begin{eqnarray}\label{ex}
\chi=\alpha\rho(1+ \beta\alpha \rho+\cdots),\\
\tilde\chi=\alpha\rho[1 +(\beta-1) \alpha\rho+\cdots].
\end{eqnarray}
(Note that with $\chi, \tilde\chi= \alpha\rho$ in the
dilute limit, $\alpha\rho$ is simply the sum of single
particle contributions, from non-interacting dipoles.)
Inserting these expansions into Eq~(\ref{87}) or
(\ref{87a}), we find, for a constant $B$-field,
\begin{equation}
{\bf f}= {\textstyle\frac{1}{2}} B^2
\mbox{\boldmath$\nabla$} [(\beta-1)(\alpha\rho)^2].
\end{equation}

There are four different microscopic models usually
employed to calculate the temperature and concentration
dependence of the susceptibility of ferrofluids: Weiss,
Onsager, mean-spherical, and high-temperature. Though
different in details, all provide the same value
$\beta=1/3$, in good agreement with experimental
data~\cite{P}.

Assuming $\chi\sim\rho$ or $\tilde\chi\sim\rho$ to hold
strictly is respectively equivalent to $\beta=0$ and
$\beta=1$, with additional restrictions for the yet
higher order terms. Both assumptions are arbitrary, and
in stark contrast to our microscopic understanding of
magnetism.

\vspace{0.5cm} {\bf Mario Liu, Institut f\"{u}r Theoretische
Physik, Universit\"{a}t Hannover, 30167 Hannover, Germany}

\end{document}